\documentstyle[12pt, righttag]{amsart}

\newtheorem{prop}{Proposition}[section]

\theoremstyle{remark}
\newtheorem{rem}{Remark}[section]

\numberwithin{equation}{section}
\begin{document}

\title[ On the integrability  of  stationary and restricted  flows ]
{On the integrability  of stationary and restricted flows of the KdV
hierarchy}

\author{G. Tondo}

\address
{G. Tondo, Dipartimento di Scienze Matematiche, Universit\`a degli Studi di
Trieste,
Piaz.le Europa 1, I34127 Trieste, Italy.}
\email{TONDO@@UNIV.TRIESTE.IT}

\date{Dipartimento di Scienze Matematiche, Universit\`a degli Studi di Trieste,
\\
Piaz.le Europa 1, I34127 Trieste, Italy.}

 \thanks
{Work partially supported by the GNFM of the Italian CNR and by the project
"Metodi Geometrici
e probabilistici in Fisica Matematica" of the Italian MURST.}

\subjclass
{Primary 58F07; Secondary 35Q58}

\maketitle

\begin{abstract}
A bi--Hamiltonian formulation for stationary flows of the KdV hierarchy is
derived in an
extended phase space.  A map between stationary flows and restricted
flows is constructed: in a case it connects an integrable H\'enon--Heiles
system and the Garnier
system. Moreover a new integrability scheme for Hamiltonian systems  is
proposed, holding
in the standard phase space.
\end{abstract}

\pagebreak

\section { Introduction}
In the last years there has been an increasing  interest for the construction
of
finite--dimensional  dynamical systems from soliton equations, through  the
so--called  methods
of {\it stationary flows}  and  {\it restricted flows} (see \cite{D},
\cite{AW1} and references
therein).  The  discovery of  suitable sets of coordinates
has allowed one to write the reduced systems as physically interesting
Hamiltonian systems.
In the case of the KdV hierarchy, the $q$--representation for  stationary flows
has given  rise to  the H\'enon--Heiles system
\cite{F, AW2},    the square eigenfunctions representation for  restricted
flows has furnished
the Neumann  and the Garnier systems \cite{C1, AW3}.
However the relation between dynamical systems which are obtained through
different
reduction techniques  from the same soliton hierarchy is not  clear; moreover
  a systematic way to find the second Hamiltonian formulation  for  stationary
flows  of any
order, without the use of a Miura map, is still lacking.
\par\noindent
The aim of this paper is to give a contribution in these directions.  In
particular:

\begin{itemize}

\item [i)]
  A bi--Hamiltonian formulation for stationary flows of the KdV
hierarchy in a suitably extended phase space is derived in a systematic way. As
an example,  the
bi--Hamiltonian structure of  H\'enon--Heiles--type systems   is explicitly
shown.

\item[ii)]
A map between stationary and restricted flows of the KdV hierarchy  is
obtained,
based on the generating function of the Gelfand--Dickey (GD) polynomials. As an
application,  a map  between  an integrable  H\'enon--Heiles system and the
Garnier system
with two degrees of freedom is constructed.

\item[iii)]
An  integrability criterion is proposed, which
can be applied   to both stationary  and   restricted flows. Though weaker than
the
bi--Hamiltonian formulation,  it does not require the extension of the phase
spaces.

\end{itemize}

The paper is organized as follows. In Sect. 2 we   construct   the
stationary flows associated to the  the KdV hierarchy through the kernel of the
 Poisson
pencil.  Using the generating function of GD polynomials as in \cite{T},   we
give a
bi--Lagrangian and a bi--Hamiltonian formulation of the Lax--Novikov
stationary equations of
any order; as an application, we exhibit a generalized H\'enon--Heiles system.
\par
In Sects. 3-4 we formulate the method of restricted flows  in terms of the
Poisson
pencil instead  of the  spectral problem as in \cite{C1,AW1}. This
formulation allows us to  explicitly construct a map between restricted and
stationary
flows,  by means of an appropriate extension of the corresponding phase spaces.
 The previous map is specialized to the H\'enon--Heiles  and the Garnier
systems.
\par
In Sect. 5 we show that  the entire
bi--Hamiltonian hierarchy  of  the H\'enon--Heiles and the  Garnier systems
cannot be  reduced
from the extended  to the standard phase space.  For this reason,   we propose
an
integrability criterion holding for a generic finite--dimensional Hamiltonian
system.  It
generalizes the criterion introduced in \cite{francesi} for the particular case
of the
H\'enon--Heiles system. Though weaker than the  bi--Hamiltonian scheme, it
  assures  Liouville--integrability of a Hamiltonian system \cite{Ar} in its
standard phase space, i.e. without the introduction of  supplementary
coordinates.   This criterion is applied to the generalized H\'enon--Heiles
system and to the
Garnier system with two degrees of freedom.
\par
Now we give some preliminaries,  mainly to fix notations
and terminology.  Let $M$ be a $n$--dimensional manifold. At any point $u \in
M$, the  tangent
and cotangent  spaces are denoted by $T_uM$ and $T_u^*M$, the pairing between
the two
spaces by
\linebreak
$<,>: T_u^*M \times T_uM \rightarrow \Bbb R$. For each smooth function $f \in
C^\infty(M)$,
 $df$ denotes the differential of $f$. $M$ is said to be a Poisson manifold if
it is endowed with a
Poisson bracket $\{ , \}: C^\infty(M) \times  C^\infty(M) \rightarrow
C^\infty(M)$, possibly
a degenerate one;
 the associated Poisson tensor $P$ is defined by $\{f,g\}(u):=<df(u),
P_u\,dg(u)>$.
So, at each point $u$, $P_u$ is a linear map $P_u: T_u^*M \rightarrow T_uM$,
skew--symmetric and with vanishing  Schouten bracket \cite{L--M}.  A function
$h \in
C^\infty(M)$ with a non trivial differential $df \in Ker P$ is called a Casimir
of $P$:
$P_u\,df(u)=0$.
 A map $\Phi:M\rightarrow M$ is a Poisson morphism if
$\{f, g\}\circ\Phi=\{f\circ \Phi, g\circ \Phi\}$, for each $f$, $g\, \in
C^\infty(M)$;  $\Phi$
leaves invariant the Poisson tensor $P$: $P_{\Phi (u)}=\Phi_*\,P_u\,\Phi^*$,
where
$\Phi_*$  and $\Phi^*$ denote, respectively,   the tangent and the cotangent
maps
associated to  $\Phi$. In particular, if the Poisson bracket is non degenerate,
i.e. if
$P$ is invertible, and the Poisson morphism is a diffeomorphism, $\Phi$ defines
 a symplectic
(canonical) transformation.  $M$ is said to be  a bi--Hamiltonian manifold
 if it is  endowed with   two  Poisson  tensors  $P_0$ and $P_1$ such that
  the associated pencil   $P^{\lambda}:= P_{1}-\lambda P_{0}$ be itself
 a Poisson tensor for any $\lambda \in \Bbb C $ \cite{ Ma1,CMP}.

\section {Stationary flows and H\'enon-Heiles systems}
\label{SectRed}

\subsection{KdV hierarchy and  Gelfand--Dickey polynomials} \label{SSectKdVGD}

Let $M$ be  a bi--Hamiltonian manifold:   if the associated Poisson pencil
$P^{\lambda}:=P_{1}-\lambda P_{0}$ admits as a Casimir  a formal Laurent series
$h(\lambda)$

\begin{equation}  \label{defCasPlambda}
h(\lambda):= \sum_{j\geq 0} h_{j}\ \lambda ^{-j} \ ,
\end{equation}
 then $h_0$ is a Casimir of  $P_{0}$ and   the coefficients $h_j \, (j\geq 1)$
are the Hamiltonian functions of  a
 hierarchy of bi--Hamiltonian vector fields $X_j$:

\begin{equation}   \label{eqXPv}
X_{j} =P_{1} dh_j= P_{0} dh_{j+1} \qquad\qquad\qquad  (j\geq 0) \ .
\end{equation}
At any point $u \in M$,  the bi--Hamiltonian flows are given by
$du/dt_j=X_{j}(u)$, $t_j$ being the evolution parameter of  the $j$th flow.
 The vector fields  \eqref{eqXPv} are Hamiltonian also with respect to the
Poisson pencil
$P^\lambda$. In fact  the recursion
relation  \eqref{eqXPv} can be written as

\begin{equation}  \label{eqXPlambdadh}
X_{j}=P^{\lambda}\, dh^{(j)} (\lambda) \ , \qquad\qquad
h^{(j)}(\lambda):=(\lambda^j h(\lambda))_{+}  \ ,
\end{equation}
where  the index
$+$ means the projection of a Laurent series onto the purely polynomial part.

 Let  $M$ be the  algebra
of polynomials in $u, u_{x}, u_{xx},\ldots$
($u=u(x)$ is a $C^\infty$ function of
$x$ and the subscript $x$ means the derivative with respect to $x$),
 and let $P_0$ and $P_1$  be
the two  Poisson tensors of the KdV hierarchy \cite{Ma1}

\begin{equation} \label{KdVP}
P_0:=\frac{d}{dx}\ , \qquad \qquad P_{1}:=\frac{d^3}{dx^3}  +4u\frac{d}{dx}
+2u_{x} \ .
\end{equation}
The gradients of the Casimirs  of the  associated Poisson pencil $P^{\lambda}$
can be obtained
searching for the 1--forms  $v(\lambda):= \sum_{j\geq 0} v_{j}\ \lambda^{-j}$
which are
solutions of the following equation

\begin{equation} \label{eqBvv=cost}
B^{\lambda}(v(\lambda),v(\lambda))=a(\lambda) \ ,
\end {equation}
where  $a(\lambda)=\sum_{j\geq -1}a_j\lambda^{-j}$,  $a_j$  are constant
parameters and
$B^{\lambda}$ is  the bilinear  function

\begin{equation} \label{defKdVBww}
B^{\lambda}(w_{1}, w_{2}):= w_{1xx} w_{2} +w_{1} w_{2xx} -w_{1x}w_{2x}  +
 (4u -\lambda) w_{1} w_{2}   \ .
\end{equation}
In fact $B^{\lambda}$   is related to the Poisson pencil through  the relation

\begin{equation} \label{eqBwwPlambda}
\frac{d}{dx} B^{\lambda}(w_{1}, w_{2}) =
 w_{1}\ P^{\lambda} w_{2} + w_{2}\ P^{\lambda} w_{1}
\qquad \qquad  (\forall\, w_1, w_2) \ .
\end{equation}
Eq. \eqref{eqBvv=cost} can  be solved developing the left hand side as a
Laurent series

\begin{equation} \label{sBvv}
B^{\lambda}(v(\lambda),v(\lambda))=\sum_{k\geq -1}B_k\lambda^{-k}
\end{equation}
 so that,
for each $a(\lambda)$, it furnishes the coefficients of the  solution
$v(\lambda)$
(unique up to a sign). The solution corresponding to
$\Bar{a}(\lambda)=-\lambda$
is the so--called basis solution $\Bar{v}(\lambda)$;   its first coefficients
are:

\begin{equation}
\Bar{v}_0=1\ , \ \Bar{v}_1=2u\  , \ \Bar{v}_2=2(u_{xx}+3u^2) \ ,
\  \Bar{v}_3=2(u^{(4)}+5u_x^2+10u_{xx}u+10u^3)
\end{equation}
and so on, namely  the gradients of the first KdV Hamiltonians. In the
following we shall consider
also the $1$--form
$v(\lambda)=c(\lambda)\Bar{v}(\lambda)$, which is solution of
\eqref{eqBvv=cost} for

\begin{equation}   \label{alambda}
a(\lambda)=-\lambda c^2(\lambda) \ , \qquad\qquad
c(\lambda)=1+ \sum_{j\geq1}c_j\lambda^{-j} \ ,
\end{equation}
 where the coefficient $c_j$ are free parameters. In this
case the first $1$--forms of the hierarchy are
$v_0=1$,  $ v_1=\Bar{v}_1+c_1$, $v_2=\Bar{v}_2+c_1\Bar{v}_1+c_2$  and so on.
\par
The coefficient $B_k$ in \eqref{sBvv} can be expressed through  the GD
polynomials. For each Laurent series
$v(\lambda)$  let us consider the  functions
$B^{(k)}(\lambda):=B^{\lambda}\left( v(\lambda),v^{(k)}(\lambda)\right)$, where
$v^{(k)}(\lambda):=\left( \lambda^k v(\lambda)\right)_{+}$;
 these functions  have the form

\begin{equation}  \label{defGDp}
B^{(k)}(\lambda)=\lambda^{k+1} v_0^2+
\sum _{j= 1}^{k-1}\lambda^{k-j}( p_{0j}-v_0 v_{j+1})+
\sum _{j\geq 0}\lambda^{-j} p_{jk} \quad (j,k\in\Bbb N_{0})\ .
\end{equation}
It can be shown that

\begin{equation} \label{p0k}
B_{-1}=-v_0^2 \ , \qquad \qquad  B_k=p_{0k}-v_0 v_{k+1}\quad (k\in\Bbb N_{0})\
;
\end{equation}
 furthermore, if  $v(\lambda)$ is a solution of Eq. \eqref{eqBvv=cost},
the coefficients $p_{jk}$ in \eqref{defGDp}  are polynomials in $u$ and its
$x$--derivatives.
They will be referred to as Gelfand--Dickey (GD)   polynomials and the function
$B^{\lambda}$
as their generating function.
\par
The {\it fundamental} property of the GD polynomials, stemming from
\eqref{defGDp},
\eqref {eqBvv=cost},
\eqref{eqBwwPlambda} and \eqref{eqXPlambdadh},  is the following relation with
the gradients
$v_j=dh_j$ and the bi--Hamiltonian vector fields  $X_k$ :

\begin{equation}       \label{eqpvX}
\frac{d}{dx} p_{jk}= v_{j} X_{k} \ .
\end{equation}
We report some GD  polynomials to be used in the following ($v_0=1$):

\begin{equation}     \label{GDp}
\begin{split}
p_{00}&=4u-v_1  \ , \\
p_{01}&=8uv_1-v_1^2-v_2+2v_{1xx} \ , \\
p_{02}&=4uv_1^2+8uv_2-2v_1v_2-v_3-v_{1x}^2+2v_1v_{1xx}+2v_{2xx} \ , \\
p_{12}&=8uv_1v_2-v_2^2+4uv_3-v_1v_3-v_4+2v_{1x}v_{2x}+2v_2v_{1xx}+2v_1v_{2xx}
+v_{3xx} \ , \\
p_{kk}&=2v_{kxx}v_k-v_{kx}^2+4uv_k^2 \ .
\end{split}
\end{equation}
The GD polynomials corresponding
to the basis solution $\Bar{v}(\lambda)$ are   the polynomials defined
in \cite[ Prop. 12.1.12]{D}.

\subsection {The method of  stationary flows} \label{SSectSflows}
The method of {\it stationary flows}  \cite{L, Nov, BN}  was
developed  in order to reduce the flows of the KdV hierarchy onto the  set
$M_n$  of  fixed
points of the $n$th flow $X_n$ of the hierarchy:

\begin{equation} \label{defMn}
M_n:=\{u\mid X_n(u,u_{x},\ldots,u^{(2n+1)})=0\} \ .
\end{equation}
 As
$M_n$ is odd-dimensional  it cannot be a symplectic manifold; nevertheless we
will  show that
it is  a bi--Hamiltonian manifold:
 it will be referred to as  {\it extended phase space}. Moreover,
$M_n$ is naturally foliated, on account of
\eqref{eqXPv} and
\eqref{KdVP}, by a one--parameter  family of $2n$--dimensional submanifolds
$S_{n}$ given
by

 \begin{equation} \label{defSn}
 S_{n}:=\{u\mid v_{n+1}(u,u_{x},\ldots,u^{(2n)})=c \}
 \end{equation}
($c$ being a constant parameter), which are  invariant manifolds  with respect
to each
vector field of the KdV hierarchy, due to the invariance of the $1$--forms
$v_k$. So  $M_n$ can
be naturally parametrized by  $v_1,\ldots, v_{n+1}$ and by
their
$x$--derivatives
$v_{1x},\ldots, v_{nx}$. We shall use these coordinates in the following.
\par
Here we perform  two different stationary reductions of the KdV flows by
improving the
procedure  introduced in  \cite{T}.  On one side,  we choose as a
reduction submanifold $S_n^{(0)}$  just the leaf $S_n$ of the  foliation
\eqref{defSn}
corresponding to $c=0$;   it is a level set of the GD polynomial
$p_{0n}$, due to \eqref{eqBvv=cost}, \eqref{sBvv} and \eqref{p0k}.  On account
of Eq.
\eqref{eqpvX}, also the GD polynomials
$p_{jn}$, restricted to
$M_n$,  are invariant with respect to each flow of the hierarchy; thus
 we can choose as a second reduction submanifold $S_n^{(1)}$  a level set of
$p_{nn}$. The
one--parameter family of the level sets of $p_{nn}$ forms a  foliation of the
manifold
$M_n$ different from the previous one. Finally we construct the bi--Hamiltonian
structure in
the ground manifold $M_n$.
\par
 From the computational point of view, one proceeds as follows.
\par\vspace{5mm}
{\bf i)}
 Due to \eqref{eqXPlambdadh} and \eqref{eqBvv=cost}, the manifold
$M_n$ is defined by the solutions $u$ of the equation

\begin{equation} \label{eqBvvn}
B^{\lambda}\big(v(\lambda),v^{(n)}(\lambda)\big)=Ê\lambda^{n} a(\lambda) \ ,
\end{equation}
where $v(\lambda)=\sum_{j=0}^{n} v_{j}\lambda^{-j}$,
$a(\lambda)=\sum_{j=-1}^{2n}
a_{j}\lambda^{-j}$.
In particular if $a(\lambda)=-\lambda c^2(\lambda)$, as in \eqref{alambda},
$M_n$ is
given by

\begin{equation} \label{LNM}
M_n=\left\{u |\ Ê\bar{X}_n+\sum_{j=1}^n c_j \bar{X}_{n-j}=0\right\} \ ,
\end{equation}
i.e. by the solutions of the Lax--Novikov equations \cite{L}.
Taking into account  Eq.~\eqref{defGDp} and
 choosing $a_{-1}=-1$, equating in Eq. \eqref{eqBvvn}
the coefficients of $\lambda^{n+1}$  we get $v_0^2=1$;  from now on we put
$v_0=1$.
Moreover  equating the coefficients of the other powers of $\lambda$  we get
the following
system:

\begin{equation} \label{sysBvvn}
p_{0k} -v_{k+1}=a_k\quad  (k=0,\ldots, n-1)\ , \qquad
p_{jn}=a_{n+j} \quad  (j=0,\ldots, n) \ .
\end{equation}
\par\vspace{5mm}
\noindent
{\bf ii)}
In order to obtain the first Poisson tensor $P_0$, we eliminate  from Eqs.
\eqref{sysBvvn} $u=v_1/2+a_0/4$ using the first equation ($k=0$) and we extract
  the   system
of $n$ second order ODE's   in the  $v_{j}\ (j=1,\ldots, n)$:

 \begin{equation}   \label{ODEnP0rsys}
 p_{0k} -v_{k+1}=a_k \quad(k=1,\ldots,n-1)\ ; \qquad p_{0n}=a_n \ ,
\end{equation}
which will be referred to as $P_{0}$--system. The remaining equations
\eqref{sysBvvn} will
furnish a  set of $n$ independent integrals of motion.
  In order to obtain a second  Poisson structure, we consider the following
system
($P_{1}$--system)

\begin{equation} \label{ODEnP1rsys}
 p_{0k}-v_{k+1} =a_k \quad (k=1,\ldots,n-1)\ ; \qquad p_{nn}=a_{2n}\ ,
\end{equation}
with $u$ as above.
\par\vspace{5mm}
\noindent
{\bf iii) }
The system (\ref{ODEnP0rsys}) can be written in Lagrangian form. To this
purpose,  we use
the so-called  Newton or $r$--representation  introduced in \cite{W2}. Namely,
we choose  as new coordinates in $S_n^{(0)}$  the first $n$ coefficients $r_j$
of
 the formal series
$r(\lambda):=\sqrt{v(\lambda)}$,

\begin{equation} \label{rcoord}
r_k=\Delta_{-k}\left( \sqrt{v(\lambda)}\right) \qquad (k=1,\ldots,n) \ ,
\end{equation}
where  $\Delta_k$ means the coefficient of $\lambda^k$ in a Laurent series.
 Taking into account  Eq.\eqref{eqBvvn},   and observing  that
$2 r_{n+1}=-\sum_{j=1}^n r_jr_{n+1-j}$, Eqs. \eqref{ODEnP0rsys} are
equivalent to

\begin{equation}
\left(\lambda^n
\big(r_{xx}+(r_1+\frac{a_0-\lambda}{4})r-\frac{a}{4r^3}\big)\right)_+=0
\ .
\end{equation}

 This system  is  Lagrangian,  with Lagrangian function

\begin{equation}
L_n^{(0)}=\Delta_{-(n+1)} \big(\cal L(\lambda; r(\lambda)\big) \ ,
\end{equation}
where $ \cal L\big(\lambda; w(\lambda)\big)$ is given,  for each Laurent series
$w(\lambda)$,
by

\begin{equation}
\cal L(\lambda; w(\lambda)):=\frac{1}{2}\big(w_x(\lambda)\big)^2-
\frac{1}{2}(w_1+\frac{a_0-\lambda}{4})w^2(\lambda)-
\frac{a(\lambda)}{8w^2(\lambda)} \ .
\end{equation}
 The Lagrangian gradients
$\frac{\delta }{\delta r_{k}}:=\frac{\partial}{\partial r_{k}}-
\frac{d}{dx}\frac{\partial}{\partial r_{kx}}$ of $L_n^{(0)}$  are

\begin{equation}
\frac{\delta L_n^{(0)}}{\delta r_{k}}=\Delta_{k-1}
\left(\lambda^n \big(-r_{xx}-(r_1+\frac{a_0-\lambda}{4})r+
\frac{a}{4r^3}\big)\right)_+
\   (k=1,\ldots,n)\ .
\end{equation}
\par
We remark that it is also
possible to put also the $P_1$--system (\ref{ODEnP1rsys}) in Lagrangian form.
To this purpose,
we take  as coordinates in $S_n^{(1)}$
$q_{k}=r_{k}\ (k=1,\ldots,n-1)$ and  $q_{n}=\sqrt{-v_{n}}$.
 By this choice the system \eqref{ODEnP1rsys} is equivalent to

\begin{equation}
\begin{split}
&\frac{1}{2}q_n^2+
\left(\lambda^{n-1}
(q_{xx}+(q_1+\frac{a_0-\lambda}{4})q-\frac{a}{4q^3}\big)\right)_+
=0  \ ,
\\  &q_{nxx}+(q_1+\frac{a_0}{4})q_n-\frac{a_{2n}}{4q_n^3}=0 \ ,
\end{split}
\end{equation}
where $\big(\lambda^{n-1} q(\lambda)\big)_+:=\big(\lambda^{n-1}
\sqrt{v(\lambda)}\big)_+$.
This is a Lagrangian system with Lagrangian

\begin{equation}
L_n^{(1)}=\Delta_{-n}\big(\cal L(\lambda; q(\lambda)\big)+
\frac{1}{2}q_{nx}^2-\frac{1}{2}(q_1+\frac{a_0}{4})q_n^2-\frac{a_{2n}}{8q_n^2} \
{}.
\end{equation}
Indeed it can be verified that the Lagrangian gradients  of $L_n^{(1)}$  are

\begin{equation}
\begin{split}
\frac{\delta L_n^{(1)}}{\delta q_{1}}&=\Delta_0
\left(\lambda^{n-1}
\big(-q_{xx}-(q_1+\frac{a_0-\lambda}{4})q+\frac{a}{4q^3}\big)\right)_+ -
\frac{1}{2}q_n^2
\\
\frac{\delta L_n^{(1)}}{\delta q_{k}}&= \Delta_{k-1} \left(
\lambda^{n-1}(-q_{xx}-\big(q_1+\frac{a_0-\lambda}{4})q+\frac{a}{4q^3}\big)\right)_+
\ (k=2,\ldots,n-1) \\
 \frac{\delta L_n^{(1)}}{\delta q_{n}}&=-
 q_{nxx}-(q_1+\frac{a_0}{4})q_n+\frac{a_{2n}}{4q_n^3}
\end{split}
\end{equation}
\par
 The two  previous Lagrangian systems can be  put in canonical Hamiltonian
form. For the
$P_0$-system the canonical momenta are $s_{n+1-k}=r_{kx} \  (k=1,\ldots ,n)$
and the
Hamiltonian function

\begin{equation} \label{Hn0}
H_n^{(0)}=\Delta_{-(n+1)}\big(\cal H(\lambda ; r(\lambda), s(\lambda))\big) \ ,
\end{equation}
where  $s(\lambda)=\sum_{j=1}^n s_j\lambda^{-j}$ and
$\cal H(\lambda ; w(\lambda), z(\lambda))$ is given by

\begin{equation}
\cal H(\lambda ; w(\lambda), z(\lambda))=\frac{1}{2}z^2(\lambda)+
\frac{1}{2}\big(w_1+\frac{a_0-\lambda}{4}\big)w^2(\lambda) +
\frac{a(\lambda)}{8w^2(\lambda)} \ .
\end{equation}
For the $P_1$--system the canonical momenta
are $ p_n=q_{nx}$, $ p_{n-k}=q_{kx}\  (k=1,\ldots, n-1)$, and the Hamiltonian
function is

\begin{equation} \label{Hn1}
H_n^{(1)}=\Delta_{-n}\big(\cal H(\lambda ; q(\lambda),p(\lambda))\big)+
\frac{1}{2}p_n^2+\frac{1}{2}(q_1+\frac{a_0}{4})q_n^2+\frac{a_{2n}}{8q_n^2}\ ,
\end{equation}
with  $p(\lambda)=\sum_{j=1}^n p_j\lambda^{-j}$.

The two Hamiltonian functions
depend, respectively,  on the two sets of coordinates and momenta
$(r_k,s_k)$, $(q_k,p_k)$  and on the two sets of free parameters
$(a_0,\ldots,a_{n-1},a_n)$ and
$(a_0,\ldots,a_{n-1}, a_{2n})$.
\par\vspace{5mm}
\noindent
{\bf iv)}
Now  let us consider the manifold $M_n$ \eqref{LNM}, which can be parametrized
either by
 $(r_k,s_k,a_n)$, or  by $(q_k,p_k,a_{2n})$, with  $a_n$ and $a_{2n}$
 as additional dynamical variables in $M_n$.  On this manifold   one can extend
trivially
the canonical Poisson structures,  the Hamiltonians and the vector fields
associated with
each one of the two systems as in \cite{AFW}. In particular the
vector fields can be extended in such a way that they are tangent to one of the
foliations
$S_{a_n}^{(0)}$ and $S_{a_{2n}}^{(1)}$.  Taking into account, on one side, the
relation between the two sets of
coordinates through the original variables
$(v_k,v_{kx})$, on the other side the relation between the two integrals of
motion $a_{n}$ and
$a_{2n}$ through the GD polynomials
$p_{0n}$ and
$p_{nn}$, a map $\Phi: M_n\rightarrow M_n, (r_k,s_k,a_{n})\mapsto
(q_k,p_k,a_{2n})$
 can be systematically constructed. It relates the Hamiltonians and the vector
fields of one
system with the corresponding ones of the other system. Since this map is not a
Poisson morphism,   the extended canonical  Poisson structures  associated with
one
chart is mapped into a  Poisson structure different from the extended canonical
structure
associated with the other chart.  If this second Poisson tensor is compatible
with the
extended canonical one, a bi--Hamiltonian formulation of the two systems is
obtained.

In conclusion we can state the following :

\begin{prop}  \label{propNSflows}
The  $P_0$--system \eqref{ODEnP0rsys} and the $P_1$--system
\eqref{ODEnP1rsys},
written respectively in the coordinates $r_k$ and $q_k$ are natural Lagrangian
systems. The
corresponding canonical Hamiltonian systems

\begin{alignat} {2}
\label{caneqHr}
r_{kx}&=\frac{\partial H_n^{(0)}}{\partial s_k} \ ,& \qquad
s_{kx}&=-\frac{\partial H_n^{(0)}}{\partial r_k} \ , \\
\label{caneqHq}
q_{kx}&=\frac{\partial H_n^{(1)}}{\partial p_k}\ ,& \qquad
p_{kx}&=-\frac{\partial H_n^{(1)}}{\partial q_k} \ ,
\end{alignat}
have $n$  integrals of motion given by

\begin{equation} \label{HK}
K_j\equiv-\frac{1}{8}p_{{jn}_{|Y}}=a_{n+j}\  (j=1,\ldots,n) \ , \quad
 H_j\equiv -\frac{1}{8} p_{{jn}_{|X}}=a_{n+j} \   (j=0,\ldots ,n-1) \ .
\end{equation}
  Moreover, the map
$\Phi : M_n\rightarrow M_n$ in the extended phase space  generates  a second
Poisson
structure.
 \end{prop}

\begin{rem}
The symbols  $|Y$ and $|X$  in \eqref{HK}  mean  that, in the GD polynomials
$p_{jk}$,  the coordinates
$(v_k,v_{kx})$ must be replaced by  the canonical coordinates   $(r_k, s_k)$
and  $(q_k,p_k)$
respectively and that   the first order $x$--derivatives  of momenta  must be
eliminated by
means of  the Hamiltonian dynamical equations \eqref{caneqHr}, \eqref{caneqHq}.
$\square$
\end{rem}
In the next Subsection we shall give some applications of the results stated in
 this
proposition.

\subsection{ The  bi--Hamiltonian structure of a H\'enon--Heiles  system}
\label{SSect2gH}

We consider  a  generalized H\'enon--Heiles  system with two degrees of
freedom.
 Its Hamiltonian is

\begin{equation} \label{2gHham}
H_0 =\frac{1}{2}\left( p_{1}^2 +p_{2}^2\right)+q_{1}^3
+\frac{1}{2}q_{1}q_{2}^2+\frac{a_4}{8q_{2}^2}
+\frac{a_0}{2}\left(q_1^2+\frac{1}{4}q_2^2\right)-\frac{a_1}{4}q_1 \ ,
\end{equation}
where $q_1,q_2,p_1,p_2$ are the canonical coordinates and momenta  and $a_0$,
$a_1$, $a_4$
are  free constant parameters.
This Hamiltonian encompasses the two cases $a_0=a_4=0$ and $a_0=a_1=0$
 introduced in \cite {B--W}.  Moreover $H_0$  is related with the Hamiltonian

\begin{equation} \label{2gfHham}
H_H=\frac{1}{2}\left( p_{1}^2 +p_{2}^2\right) + \frac{1}{2}\left( A q_1^{'2}+B
q_2^{'2}\right) +
q_{1}^{'3} +\frac{1}{2}q'_{1}q_{2}^{'2} +\frac{a_4}{8q_{2}^{'2}} \ ,
\end{equation}
through the map

\begin{equation} \label{mapHHf}
 q_1=q_1'+ \frac{A}{2}-2B \ ,\   q_2=q_2' \ ,\  a_0=-2A+12B\ , \  a_1=-A^2+16
AB-48 B^2 .
\end{equation}
The function $H_H$ is the Hamiltonian of a
 classical integrable H\'enon--Heiles system \cite{Ta} with the
additional term $a_4/8q_2^{'2}$.
\par
The function \eqref{2gHham} is the Hamiltonian of the
 the  vector field obtained reducing $X_0(u)=u_x$   to the stationary manifold
$M_2$
given by the fixed points  of the flow $X_2+c_1 X_1+c_2 X_0$

\begin{equation}
M_2=
\left\{u| u^{(5)}+10 u_{xxx}u+20u_{xx}u_x+30u_xu^2+c_1(u_{xxx}+6u_xu)+c_2
u_x=0\right\}
\ ,
\end{equation}
where $c_1=-a_0/2$, $c_2=-a_1/2+a_0^2/4$.
\par
It can be obtained specializing to the case $n=2$ the Hamiltonian \eqref{Hn1}
of the
$P_1$--system. In this case $H_2^{(1)}=H_0$ and the canonical coordinates  and
momenta are, respectively,
 $q_{1}=v_1/2, q_{2}=\sqrt{-v_2}$, $p_{1}=q_{1x}$, $ p_{2}=q_{2x}$. The
 integrals of motion   obtained by the reduction of the GD polynomials are

  \begin{equation} \label{H2H}
\begin{split}
H_{0}&\equiv-\frac{1}{8} p_{{02}_{| X}} \ , \qquad
H_{2}\equiv-\frac{1}{8} p_{{22}_{| X}}=-\frac{a_4}{8} \ , \\
 H_{1}& \equiv-\frac{1}{8}p_{{12}_{| X}}=p_{2}^2 q_{1} -p_{1}p_{2}q_{2}
        -\frac{1}{2}q_{1}^2 q_{2}^2 -\frac {1}{8}q_{2}^4
        + \frac{a_4q_{1}}{4q_{2}^2} -\frac{a_0}{4}q_1q_2^2+\frac{a_1}{8}q_2^2
\ .
\end{split}
 \end{equation}
The  corresponding Hamiltonian vector fields will be denoted by  $X_{j+1}:= E\,
dH_j\ (j=0,1, 2)$;
$E$ being the
 canonical $(4\times 4)$ Poisson  matrix. The  H\'enon--Heiles vector field
$X_1$ is:

\begin{equation} \label{2gHX0}
X_1=[ p_1, p_2,-3q_1^2-\frac{1}{2}q_2^2-a_0q_1+\frac{a_1}{4} ,
-q_1q_2+\frac{a}{4q_2^3}-\frac{a_0}{4} q_2]^T .
\end{equation}
The second Hamiltonian formulation can be obtained specializing to the case
$n=2$ the Hamiltonian \eqref{Hn0} of the $P_0$--system:

\begin{equation}
H_2^{(0)}=s_1s_2-\frac{5}{8}r_{1}^4
+\frac{5}{2}r_{1}^2r_{2}-\frac{1}{2}r_{2}^2
-\frac{a_0}{2}r_1^3+\frac{3}{8}a_1r_1^2+a_0r_1r_2
-\frac{a_2}{4}r_1-\frac{a_1}{4}r_2 \ ,
\end{equation}
where  the canonical coordinates
\eqref{rcoord} and  momenta are, respectively,
$r_1=v_1/2$, $r_2=v_2/2-v_1^2/4$, $s_{1}=r_{2x}$, $s_{2}=r_{1x}$. The
 integrals of motion   obtained by the reduction of the GD polynomials are

\begin{equation} \label{K2H}
\begin{split}
K_0 &\equiv -\frac{1}{8}  p_{{02}_{| Y}}= -\frac{a_2}{8} \ , \qquad
K_1\equiv-\frac{1}{8} p_{{12}_{|Y}}=H_2^{(0)}\ , \\
K_2& \equiv-\frac{1}{8} p_{{22}_{| Y}}= -
s_2^2r_2+s_1s_2r_1+\frac{1}{2}s_1^2-\frac{1}{2}r_1^5
+2r_1r_2^2-\frac{3}{8}a_0r_1^4 +\\
&+\frac{a_1}{4}r_1^3-\frac{a_0}{2}r_1^2r_2+\frac{a_1}{2}r_1r_2+\frac{a_0}{2}r_2^2
-\frac{a_2}{8}r_1^2 -\frac{a_2}{4}r_2 \ ,
\end{split}
\end{equation}
and the corresponding   Hamiltonian vector fields will be denoted by  $Y_j:=
E\, dK_j$.
\par\par
Now we construct the bi--Hamiltonian structure of the H\'enon--Heiles system.
Let $M_2$ be the
$5$--dimensional extended phase space parametrized by
$(r_1,r_2, s_1,s_2; a_2)$ or $  (q_1, q_2, p_1, p_2; a_4)$.
It is convenient to make use of   block notations. So, for example, we
denote with
$(r, s; a)$ the $5$--tuple $(r_1,r_2,s_1,s_2; a_2)$, with
$\Tilde X=[\Tilde X^r, \Tilde X^s; \Tilde X^a]^T$   the generic  vector field
and  with
$d\Tilde K=[ \partial \Tilde K /\partial r , \partial \Tilde K /\partial s ;
\partial \Tilde K/\partial a ]^T$
the generic gradient of a function $\Tilde K$ (the superscript $T$ means
transposition). In this
notation a  vector field
$\Tilde X=\Tilde P\, d\Tilde K$
  with Hamiltonian function $\Tilde K$ with respect to a Poisson tensor $\Tilde
P$ will
be written
\begin{equation}
 \begin{bmatrix}
\Tilde X^r \\\
 \Tilde X^s\\\
 \Tilde X^a
\end{bmatrix}
=\begin{bmatrix}
P^{rr} &P^{rs}&P^{ra} \\\
P^{sr} &P^{ss}&P^{sa} \\\
P^{ar}&P^{as}&P^{aa}
\end{bmatrix}
  \begin{bmatrix}
\frac{\partial \Tilde K }{\partial r}\\\
 \frac{\partial \Tilde K }{\partial s}  \\\
\frac{\partial \Tilde K}{\partial a}
\end{bmatrix} \ ,
\end{equation}
where $P^{sr}=-{(P^{rs})}^T$ etc. $\ldots$.
{}From the definition of $r_1,r_2$ and $q_1, q_2$ in terms of $v_1$ and $v_2$,
and from
\eqref{H2H} and \eqref{K2H} one obtains  the following map
$\Phi : M_2\rightarrow M_2, (r, s ; a_2) \mapsto (q, p; a_4)$

\begin{alignat}{3}   \label{map2HP0P1}
q_1&=r_1\ ,\quad&\quad q_2&=(-2r_2-r_1^2)^{1/2}
    &   &\notag\\
p_1&=s_2\ ,\quad&\quad p_2&=-\frac{s_1+r_1s_2}{(-2r_2-r_1^2)^{1/2}}\ , \qquad
&a_4&=- 8 K_2
\end{alignat}
with $K_2$ given by  Eq.\eqref{K2H}. In these two charts let us consider the
extended
Hamiltonians
$\Tilde{H}_j$ and
$\Tilde{K}_j$,  the vector fields
$\Tilde{X}_j$ ($\Tilde{X}_j^r=X_j^r$, $\Tilde{X}_j^s=X_j^s$, $\Tilde{X}_j^a=0$)
and
$\Tilde{Y}_j$ ($\Tilde{Y}_j^r=Y_j^r$, $\Tilde{Y}_j^s=Y_j^s$,
$\Tilde{Y}_j^a=0$),
the  extension of the canonical Poisson structure,
$\Tilde{E}:= \bigl (
 \begin{smallmatrix}
 {\boldsymbol 0}& {\boldsymbol1}& 0\\
-{\boldsymbol1}&{\boldsymbol 0} & 0 \\
  0& 0& 0
\end{smallmatrix}
 \bigr )  $. The following proposition holds

\begin{prop}
The action of the map $\Phi : M_2 \rightarrow  M_2$ defined by
\eqref{map2HP0P1}   on the
Hamiltonians $\Tilde{H}_j$, the vector fields $\Tilde Y_j$  and   the  Poisson
tensor
$\Tilde{P}'_0:=\Tilde E$ is given by $\Phi ^* (\Tilde{H}_j)=\Tilde{K}_j $,
$\Phi_*(\Tilde{Y}_j)=\Tilde{X}_{j}$ and by

\begin{equation}
\Tilde{P}_0 :=\Phi _* \Tilde{P}'_0 \Phi ^* =
\begin{bmatrix} \label{2gHexP0}
                            0&                             A& -8\Tilde X_{2}^q
\\\
                       -A^T&                             B& -8 \Tilde X_{2}^p
\\\
 8(\Tilde X_{2}^q)^T&  8(\Tilde X_{2}^p)^T&0
\end{bmatrix} \ ,
\end{equation}
where
$ A=\frac{1}{q_2^2}\bigl (
\begin{smallmatrix}
0&-q_2 \\
-q_2&2q_1
\end{smallmatrix} \bigr ) $,
$ B=\frac{1}{q_2^2}\bigl (
\begin{smallmatrix}
0&-p_2  \\
p_2 &0
\end{smallmatrix} \bigr ) .
\square$
\end{prop}

Thus we have recovered in the extended phase space $M_2$ a second  Poisson
tensor
  $\Tilde P_0$. We can check that   $\Tilde P_0$ is compatible with
$\Tilde{P}_1=\Tilde{E}$.
Furthermore $\Tilde{P}_0$ and $\Tilde{P}_1$ give rise to the following
bi--Hamiltonian
hierarchy

\begin{equation} \label{2gHX}
\Tilde{X}_{j+1}:=\Tilde{P}_1\, d\Tilde{H}_{j}=\Tilde P_0\, d\Tilde{H}_{j+1}
\qquad (j=0,1)\ ,
\end{equation}
the Hamiltonians $\Tilde H_0$ and $\Tilde H_2$ being Casimirs of $\Tilde P_0$
and
$\Tilde{P}_1$ respectively.

\section { Restricted flows and  Garnier systems} \label{SSectRflows}

The method of restricted  flows was introduced in \cite{Mo}  as a
{\it non linearization} of the  KdV spectral problem and was generalized in
 \cite{C1, AW1}.
We formulate this method putting the emphasis on the role  of the GD
polynomials and of
their generating function; this formulation allows us to
 construct a map between stationary and restricted flows in the next section.
In view of the
applications, we begin by applying the method to the KdV hierarchy,  recovering
the Garnier
system.
\par
Let us consider the following system

\begin{equation}   \label{eqSER}
p_{00}-v_1={a}_0, \quad
P_0 \big(v_{1} -\sum_{j=1}^n\beta_{j}\big)=0 ,\quad
 P^{\lambda_k} \beta_k =0 \quad  (k= 1,\ldots,n)
\end{equation}
where: $\lambda_1,\ldots,\lambda_n$ are  distinct fixed parameters,
$P^{\lambda_k}:=P_1 -\lambda_k P_0$  ($P_0$ and $P_1$ being the two KdV Poisson
tensors    \eqref{KdVP}). This is a  system of $(n+2)$
equations in  $u, v_1, \beta_1,\ldots,\beta_n$. The  second equation will be
referred
to as the $P_0$--{\it restriction} of the first  KdV flow ${X}_0=P_0 v_1=
v_{1x}$,
and the last
$n$ equations  define  the kernel of $n$ Poisson tensors extracted from the
Poisson pencil.
On account of \eqref{GDp}, \eqref{KdVP} and\eqref{eqBwwPlambda}  this  system
is equivalent
to the following one

\begin{equation}  \label{eqBarg}
u=\frac{v_1}{2} +\frac{a_0}{4}\ , \qquad v_1=\sum_{j=1}^n\beta_{j}+c \ ,
\qquad B^{\lambda_k}(\beta_k,\beta_k)=f_k \ ,
\end{equation}
where $c$  and $f_k$ are free
parameters and $B^{\lambda}$ is just
the generating function  \eqref{defKdVBww} of the GD polynomials.

 Using the first two equations  to eliminate $u$ and $v_1$ from the last  $n$
equations,  one gets a system of $n$ ODE's of second order  for
 $\beta_1,\ldots,\beta_n$: .
\begin{equation}  \label{eqBbb}
  2 \beta_{kxx}\beta_k -\beta_{kx}^2 +2\beta_k^2(\sum_{j=1}^n \beta_j +d)
-\lambda_k
\beta_k^2 =f_k  \quad  (k=1,\ldots,n)  \ ,
 \end{equation}
where $d:=c+a_0/2$. Introducing the so--called eigenfunction variables
$\psi_j=\sqrt{\beta_j}$
and the  momenta
$\chi_j =\psi_{jx}$, Eqs.\eqref{eqBbb} can be written in canonical Hamiltonian
form

\begin{equation} \label{caneqG}
\psi_{jx}=\frac{\partial {\cal K}_G}{\partial\chi_j}\ , \qquad\qquad
\chi_{jx}=-\frac{\partial {\cal K}_G}{\partial\psi_j}\qquad (j=1,\ldots ,n) \ .
\end{equation}

with Hamiltonian

\begin{equation} \label{nGham}
{\cal K}_G=\frac{1}{2} \sum_{j=1}^n \chi_j^2+
\frac{1}{8}\left[\left(\sum_{k=1}^n
\psi_j^2\right)^2
  -\sum_{j=1}^n (\lambda_j-2d)\psi_j^2 + \sum_{j=1}^n \frac{f_j}{\psi_j^2}
\right] \ .
 \end{equation}
The corresponding Hamiltonian vector field ${\cal  Y}_G =\cal  E\, d{\cal
K}_G$ is

\begin{equation} \label{2gGY1}
\cal  Y_G=[\chi_j,
 -\frac{1}{2}(\psi_1^2+\psi_2^2)\psi_j+\frac{1}{4}(\lambda_j-2d)\psi_j
+\frac{f_j}{4\psi_j^3}]^T
\   (j=1,\ldots ,n)  \ ,
\end{equation}
 $\cal  E$ being the $(2n\times 2n)$ canonical Poisson matrix. Eqs.
\eqref{caneqG} are just the
equations of  the Garnier system with $n$ degrees of freedom
\cite{AW1}. A set of  integrals of motion is

\begin{equation} \label{nGI}
I_j=\chi_j^2+
\frac{\psi_j^2}{4}\big(2d-\lambda_j+\sum_{k=1}^n\psi_k^2\big)+
\frac{f_j}{4\psi_j^2}+
\sum^n \begin{Sb} k=1 \\\  k\not=j \end{Sb}
\frac{1}{4\lambda_{jk}}
\big(\frac{f_j\psi_k^2}{\psi_j^2}+\frac{f_k\psi_j^2}{\psi_k^2}+
(\psi_j\chi_k-\psi_k\chi_j)^2\big) \ ,
\end{equation}
with   $\sum_{j=1}^n I_j=2{\cal K}_G$. These integrals
were obtained  in \cite{W1} by means of a Lax representation; we shall recover
them in the
next section by the use of the generating function of the GD polynomials.
\par
 Let us consider   the $(2n+1)$ extended phase space $\cal M_2$ with
coordinates
$(\psi_k,\chi_k;d)$ and the extended Hamiltonian $\Tilde{\cal  K}_G$,
the  vector field $\Tilde{\cal  Y}_G =\Tilde{\cal  E}\, d\Tilde{\cal  K}_G$
with
$\Tilde{\cal  E}= \bigl (
 \begin{smallmatrix}
 {\boldsymbol 0_n}& {\boldsymbol1_n}& 0\\
-{\boldsymbol1_n}&{\boldsymbol 0_n } & 0 \\
  0& 0& 0
\end{smallmatrix}
\bigr ) $. In this space the Garnier system has a second Hamiltonian structure
  given by

\begin{equation}  \label{nGP1}
\Tilde{\cal  P}_1 :=
\begin{bmatrix}   \label{2gGexP0}
0&\Lambda-\psi\otimes\psi&4\Tilde{\cal Y}_{G}^\psi\\\
-(\Lambda-\psi\otimes\psi)^T&\chi\otimes\psi-\psi\otimes\chi&4\Tilde{\cal
Y}_{G}^\chi \\\
 -4(\Tilde{\cal Y_{G}}^\psi) ^T&-4(\Tilde{\cal Y_{G}}^\chi )^T&0  \\\
\end{bmatrix} \ ,
\end{equation}
 where $\otimes$ denotes the tensor product,
$\psi=[\psi_1,\ldots,\psi_n]^T$, $\chi=[\chi_1,\ldots,\chi_n]^T$,
 $\Lambda=diag(\lambda_1,\ldots,\lambda_n)$. This structure is an  extension of
the
one constructed in \cite{AW3} for $f_k=0\ , (k=1,\ldots, n)$.
In view of the applications we specialize the  above structure to the case
$n=2$, in
 the five--dimensional extended phase space $\cal M_2$ with coordinates
$(\psi_1,\psi_2,\chi_1,\chi_2;d)$. The following proposition holds:

\begin{prop}
The  Garnier vector field $\Tilde{\cal  Y}_1=\Tilde{\cal  Y}_G$ belongs to the
following
bi--Hamiltonian hierarchy

\begin{equation} \label{2gGY}
\Tilde{\cal  Y}_{j+1}=\Tilde{\cal  P}_1\, d\Tilde{\cal  G}_j=
\Tilde{\cal  P}_0\, d\Tilde{\cal G}_{j+1}
\qquad (j=0,1)\ ,
\end{equation}
where the Hamiltonians $\Tilde {\cal  G}_j$ are given by
\begin{equation} \label{2gGG}
\begin{split}
\Tilde{\cal  G}_0&=\frac{d}{4} \ , \qquad
\Tilde {\cal  G}_1=-(\lambda_1+\lambda_2)\frac{d}{4}+\frac{1}{2}(\Tilde
I_1+\Tilde I_2) \\
\Tilde{\cal  G}_2&=\lambda_1 \lambda_2\frac{d}{4}
-\frac{1}{2}(\lambda_1+\lambda_2)(\Tilde I_1+\Tilde I_2)+
\frac{1}{2}(\lambda_1\Tilde I_1+\lambda_2\Tilde I_2)
 \ ,
\end{split}
\end{equation}
 $\Tilde {\cal  G}_0$ and $\Tilde{\cal  G}_2$ being Casimirs of $\Tilde{\cal
P}_0$ and
$\Tilde{\cal  P}_1$ respectively, and $\Tilde {I}_1, \Tilde I_2$ being the
extensions to
$\cal M_2$ of the integrals of motion \eqref{nGI}.
\end{prop}

As in the case of the H\'enon--Heiles system, a
bi--Hamiltonian structure for the Garnier system seems to naturally exist only
in its
extended phase space. Nevertheless  in Subsect. \ref{SSect2gHGintstr} a
realization of
the  integrability structure introduced in  Prop. \ref{TeoC} will be
constructed  in the
original four--dimensional phase space.

\section{ A map between  stationary  and  restricted flows}
\label{SSectmap}

Now we shall construct a map between  the $n$th stationary
flow and  the previous restricted flow of the KdV hierarchy. To this end we
extend the
corresponding phase spaces, regarding some free parameters  in the Hamiltonian
functions as
additional dynamical variables.

\subsection{The general case}
As for  the $P_1$--formulation of the stationary  flow
\eqref{caneqHq} we
  extend its phase space to a $(3n+1)$--dimensional space, $\Tilde M_n$, with
coordinates
 $( q_k,p_k; a_0,\ldots ,a_{n-1}, a_{2n})$; analogously we consider
 the $P_0$--formulation of the first restricted flow \eqref{caneqG} in the
extended space
$\Tilde{\cal M}_n$ with coordinates  $(\psi_k,\chi_k; f_1,\ldots,f_n, d)$.
\par
 Let us consider the solutions   $q_k$  of the dynamical equations
\eqref{caneqHq};  then  $v^{(n)}(\lambda)$ given by

\begin{equation}
v^{(n)}(\lambda)=\lambda \left(q^2(\lambda)\right)^{(n-1)}-q_n^2 \ ,
\end{equation}
with $q(\lambda)=1+\sum_{j=1}^nq_j\lambda^{-j} $, satisfies \eqref{eqBvvn} and
consequently
the following equation

\begin{equation} \label{Bstaz}
B^{\lambda} \big(v^{(n)}(\lambda),v^{(n)}(\lambda)\big)=\lambda^{2n}a(\lambda)
 \ ,
\end{equation}
where, as above, we put $u=v_1/2+ a_0/4$. So,  for each $n$--tuple of
distinct complex parameters
$\lambda_j$,   any solution $ v^{(n)}(\lambda)$ of Eq. \eqref{Bstaz} fulfills
the
 system

\begin{equation} \label{Blambdakvnvn}
B^{\lambda_k} \big(v^{(n)}(\lambda_k),
v^{(n)}(\lambda_k)\big)=\lambda_k^{2n}a(\lambda_k)  \ \  (k=1,\ldots,n) \ ,
\end{equation}
where
$v^{(n)}(\lambda_k):={v^{(n)}(\lambda)}_{|\lambda=\lambda_k}$.
In order to have a solution   satisfying also
the second  equation \eqref{eqBarg},  the  Lagrange interpolation formula can
be used
\cite{Al, C2}. It allows us to represent  the polynomial $v^{(n)}(\lambda)$  by

\begin{equation} \label{vinterpol}
v^{(n)}(\lambda)= p(\lambda) +
\sum_{j=1}^n\frac{p(\lambda)}{\lambda-\lambda_j}\beta_j \ ,
\end{equation}
where $p(\lambda)=\prod _{j=1}^n(\lambda-\lambda_j)$, and

\begin{equation} \label{betak}
 \beta_k= \frac{v^{(n)}(\lambda_k)}{p'(\lambda_k)} \qquad\qquad (k=1,\ldots,n)
\ ,
\end{equation}
( $p'(\lambda)$ means the derivative of $p(\lambda)$ with respect to
$\lambda$).
 Obviously  the $n$ functions $\beta_k$ \eqref{betak} are solutions of the
following
system

\begin{equation}  \label{eqBbba}
  2 \beta_{kxx}\beta_k -\beta_{kx}^2 +2\beta_k^2(\sum_{j=1}^n \beta_j
+\frac{a_0}{2}-\sum_{j=1}^n \lambda_j)-\lambda_k
\beta_k^2 =\frac{\lambda_k^{2n}  a(\lambda_k)} {(p'(\lambda_k))^2} \qquad
(k=1,\ldots,n)
\ ;
 \end{equation}
furthermore  $\beta_k$ satisfy the so--called Bargmann constraint

\begin{equation}
\sum_{j=1}^n \left( \beta_j- \lambda_j \right)=v_1 \ ,
\end{equation}
as one can  verify by means of \eqref{vinterpol}.
Comparing \eqref{eqBbba} with \eqref{eqBbb}, we can state the following

\begin{prop}
Let   $\Psi : \Tilde M_{n}\rightarrow
\Tilde{\cal M}_n,   (q, p; a_0,\ldots ,a_{n-1}, a_{2n})\mapsto
(\psi,\chi; f_1,\ldots,f_n, d)$ be the map:

 \begin{equation} \label{coormapHG}
\begin{split}
\psi_k& =
\big(\frac{\sum_{j=0}^{n-1}\sum_{l=0}^jq_lq_{j-l}\, \lambda_k^{n-j}-q_n^2}
{p'(\lambda_k)}\big)^{1/2}
\ , \quad
\chi_k =\frac{\sum_{j=1}^{n-1}\sum_{l=1}^j q_{j-l}p_{n-l}\,
\lambda_k^{n-j}-q_np_n}
      {\left(p'(\lambda_k)\big(
\sum_{j=0}^{n-1}\sum_{l=0}^jq_lq_{j-l}\,
\lambda_k^{n-j}-q_n^2\big)\right)^{1/2}} \ ,\\
 f_{k} &= \frac{1}{(p'(\lambda_k))^2}\big(a_{2n}-8\sum_{j=0}^n
H_{n-j}\,\lambda_k^j
+\sum_{j=n+1}^{2n+1} a_{2n-j}\,\lambda_k^j\big) \ , \quad
d=\frac{a_0}{2}-\sum_{j=1}^n\lambda_j
\end{split}
\end{equation}
$(k=1,\ldots,n)$, where $H_j$ are the Hamiltonian functions \eqref{HK}. If
$(q_k,p_k)$ are solutions of the stationary flows \eqref{caneqHq}, then
$(\psi_k,
\chi_k)$ are solutions of the Garnier system \eqref{caneqG}
for $f_k$ and $d$ given by \eqref{coormapHG}.
\end{prop}

\begin{rem}
The function $B^{\lambda}$ is  also a generating function of  integrals of
motion
for the Garnier system.  Indeed evaluating the function $B^{\lambda}$ by means
of
 \eqref{vinterpol} and eliminating the first
$x$--derivatives of $\chi_k$ by means of the Hamilton equations \eqref{caneqG},
one gets

\begin{equation} \label{eqInG}
4\sum_{j=1}^n\frac{I_j}{\lambda-\lambda_j}+
\sum_{j=1}^n\frac{f_j}{(\lambda-\lambda_j)^2}+2d-\lambda=
\frac{\lambda^{2n}\hat a(\lambda)} {(p(\lambda))^2}
\end{equation}
where $I_j$ are the functions \eqref{nGI}.
 Taking in this equation the residues at $\lambda=\lambda_j$ it follows that
the functions $I_j$
are integrals of motion along the flow \eqref{caneqG}. $\square$
\end{rem}

\subsection{ The map between the H\'enon--Heiles  and the Garnier system}

Now we specialize the   map  of Prop. 4.1 to the
H\'enon--Heiles  and the Garnier systems with two degrees of freedom:  we
obtain the
surprising result that the H\'enon--Heiles vector field is mapped into the
Garnier vector field.
 Let us   consider the seven--dimensional phase space of
the H\'enon--Heiles system  $\Tilde M_2$ with coordinates
$(q, p; a_0,a_1,a_{4})$. Similarly, for the Garnier systems let us select the
parameters
$f_{1},f_{2},d$ and
enlarge the phase space to a seven--dimensional phase space $\Tilde{\cal
M}_2$, with
coordinates $(\psi,\chi; f_{1},f_{2},d)$. It is easy to prove the following

\begin{prop}  \label{propmapHHG}
 Let  $\Psi :\Tilde M_2\rightarrow \Tilde{\cal  M}_2,
(q, p; a_0,a_1,a_{4})\mapsto  (\psi,\chi; f_{1},f_{2},d)$ be  defined by

\begin{equation}   \label{map2HG}
\begin{split}
\psi_{1} &= \lambda_{12}^{-1/2}(\lambda_{1}^2 +2\lambda_{1}q_{1}
-q_{2}^2)^{1/2}
 \ ,\qquad\qquad
\psi_{2} = \lambda_{12}^{-1/2} (-\lambda_{2}^2-2\lambda_{2}q_{1}
+q_{2}^2)^{1/2} \ , \\
\chi_{1}& = \frac {\left( \lambda_{1}p_{1} -q_{2}p_{2}\right)}
{\big(\lambda_{12}\ (\lambda_{1}^2
                 +2\lambda_{1}q_{1} -q_{2}^2)\big)^{1/2}} \ , \qquad
\chi_{2} = \frac {\left( \lambda_{2}p_{1} -q_{2}p_{2}\right)}
{\big(\lambda_{12}\ (-\lambda_{2}^2
                 -2\lambda_{2}q_{1} +q_{2}^2)\big)^{1/2}}  \ , \\
f_{1} &=  \lambda_{12}^{-2} \left(-\lambda_{1}^5+a_0\lambda_1^4+a_1\lambda_1^3
-
               8 H_{0}\lambda_{1}^2 -8 H_{1}\lambda_{1} +a_4 \right) \ , \\
f_{2} &= \lambda_{12}^{-2} \left(-\lambda_{2}^5+a_0\lambda_2^4+a_1\lambda_2^3 -
            8 H_{0}\lambda_{2}^2 -8 H_{1}\lambda_{2} +a_4 \right) \ ,
\quad
 d=\frac{a_0}{2}-(\lambda_1+\lambda_2)\ ,
\end{split}
\end{equation}
where $\lambda_{12}=\lambda_1-\lambda_2$.
The tangent map $\Psi_*$ maps the extended H\'enon-Heiles vector fields
$\Tilde{X_{1}},
\Tilde{X_{2}}$ \eqref{2gHX} into the extended Garnier vector fields
$\Tilde{\cal  Y}_{1}, \Tilde{\cal Y}_{2}$ \eqref{2gGY}:

\begin{equation} \label{mapXH}
 \Psi_{*}\left(\Tilde{X_{1}}\right) = \Tilde{\cal  Y_{1} }\ , \qquad\qquad
\Psi_{*}\left(\Tilde{ X_{2}}\right) =\Tilde{\cal  Y_{2}} \ .
\end{equation}
Moreover the pull--back of the Garnier integrals of  motion $\cal  G_{1}$ and
$\cal  G_{2}$ are integrals of motion for the H\'enon--Heiles system
\begin{equation}
\begin{split}
\Psi^{*} (\cal  G_{1})&=-\frac{1}{8}
(\lambda_1^2+\lambda_2^2)+\frac{a_0}{8}(\lambda_1+\lambda_2)+\frac{a_1}{8}  \\
\Psi^{*} (\cal  G_{2}) &=\lambda_{12}^{-2}
              \bigl(  2
\lambda_{1}\lambda_{2}H_0+(\lambda_{1}+\lambda_{2})H_1+2H_2\big) +\\
       &+\frac{ \lambda_{12}^{-2}\lambda_1\lambda_2}{4}
 \left((\lambda_{1}^3+\lambda_{2}^3 )
-\frac{a_0}{2}(\lambda_{1}^2+\lambda_{2}^2)
-\frac{a_1}{2}(\lambda_1+\lambda_2) \right)  \ ,
\end{split}
\end{equation}
The action of the map $\Psi$  on
the Poisson tensor
$\Tilde E$ of the H\'enon--Heiles system,  furnishes a new Poisson tensor for
the Garnier system
compatible with $\Tilde{\cal E}$. Moreover the  action of
$\Psi$  on the Poisson tensor $\Tilde P_0$  is given by

\begin{equation}
\Psi^*\Tilde P_0\Psi_*=\lambda_{12}^{-2}
\begin{bmatrix}
   0&\cal A&0 \\\
-\cal A^T&\cal B&0\\\
 0&0&0
\end{bmatrix} \ ,
\end{equation}
where
\begin{equation}
\begin{split}
\cal A&=\frac{1}{\psi_1^2\psi_2^{2}}
\begin{bmatrix}
\psi_2^2(\psi_1^2+\psi_2^2+\lambda_1-\lambda_2)&
-\psi_1\psi_2(\psi_1^2+\psi_2^2)  \\\
-\psi_1\psi_2(\psi_1^2+\psi_2^2) &
\psi_1^2(\psi_1^2+\psi_2^2+\lambda_2-\lambda_1)
\end{bmatrix}  \ , \\
\cal B&=\frac{\psi_1^2+\psi_2^2}{\psi_1^2\psi_2^{2}} \begin{bmatrix}
0&(\chi_2\psi_1-\chi_1\psi_2)\\\
-(\chi_2\psi_1-\chi_1\psi_2) &0
\end{bmatrix} \ .
\end{split}
\end{equation}
\end{prop}
 So the map $\Psi$ is not a Poisson morphism. However, according to Eqs.
\eqref{mapXH}, the
orbits of the H\'enon--Heiles system are mapped   into the orbits of the
Garnier
system.

\section{A new integrability structure}

\subsection{The reduced  structures of the  H\'enon--Heiles  and   the Garnier
systems}
\label{SSect2gHredstr}

In order to have a bi--Hamiltonian hierarchy  also in the original phase space
 for the  H\'enon--Heiles and the Garnier systems, one can try to apply the
reduction techniques
known
from the literature \cite{L--M, MR}. In particular, two methods can
be followed:   a {\it restriction} to the standard phase space or a
{\it projection} onto it.  However, in both cases, these attempts fail.
\par
As for the H\'enon--Heiles system,  if
the restriction submanifold is chosen to be a leaf $S_{a_4}^{(1)}$ of the
second natural
foliation in $M_2$,  the Hamiltonians $\Tilde{H}_j$, the vector fields
$\Tilde{X}_j$ and the
Poisson structure
$\Tilde{P}_1$ can  be trivially restricted respectively to $H_j$, $X_j$ and
$E$; but it turns out
that  $\Tilde P_0$ cannot be restricted.  So   two integrable Hamiltonian
vector fields   are
obtained in $S_{a_4}^{(1)}$ but  not a bi--Hamiltonian hierarchy.
 \par
If $\Pi: M_2 \rightarrow S_2 ,(q_1,q_2,p_1,p_2;a_{2n})\mapsto
(q_1,q_2,p_1,p_2)$ is the projection map, the  Hamiltonians $\Tilde{H}_j$ and
the
vector fields $\Tilde{X}_j$  cannot be projected onto $S_2$, because they
depend on the fiber
coordinate. Instead, the Poisson tensors
$\Tilde{P}_0$  and $\Tilde{P}_1$ are projected onto:

\begin{equation} \label{2gHP0}
P_H:=\Pi _* \Tilde P_0 \Pi ^* =
\begin{bmatrix}
0&A \\\
-A^T&B
\end{bmatrix}  \ ,
\qquad\qquad
 \Pi _* \Tilde{P}_1 \Pi ^*=E \ ,
\end{equation}
with $A$, $B$ as in Prop. 2.2. Because these operators are compatible and
invertible,   one
obtains the following Nijenhuis tensor \cite{M}

\begin{equation} \label{2gHN}
N_H:= P_H E^{-1}=
\begin{bmatrix}
A&0 \\\
B&A^ T
\end{bmatrix}
\end{equation}
and consequently  the  hierarchy of Poisson tensors
$P_k:=N_H^k P_H\ ,  k \in {\Bbb Z }$.
However these tensors  are not invariant along the flow of the H\'enon--Heiles
vector
field $X_1$ \eqref{2gHX0}.
In other words $X_1$ is neither a symmetry of $P_0$ nor of $P_1$, so that these
tensors cannot
generate a bi--Hamiltonian hierarchy starting from $X_1$.
\par
As in the case of the H\'enon--Heiles system, one cannot   reduce the
bi--Hamiltonian
structure of the Garnier system with $n$ degrees of freedom  onto the
restricted phase space.
 If $\Pi:\cal  M_n\rightarrow \cal  S_n^{(1)} ,
(\psi_k,\chi_k; d) \mapsto (\psi_k,\chi_k)$ is the projection map,
the Poisson tensor $\Tilde {\cal  P}_0$ and $\Tilde {\cal  P}_1$ are projected
onto  two
compatible tensors

\begin{equation}
\Pi _* \Tilde{\cal  P}_0 \Pi ^*=\cal  E \ , \qquad\qquad
\cal P_G:=\Pi _* \Tilde {\cal  P}_1 \Pi ^* =
\begin{bmatrix}
0&\Lambda-\psi\otimes\psi \\\
-(\Lambda-\psi\otimes\psi)^T&\chi\otimes\psi-\psi\otimes\chi \\\
\end{bmatrix}  \ .
\end{equation}
They   give rise to the  Nijenhuis tensor
$\cal  N_G:=\cal  P_{G}\cal  E^{-1}$
together with the  hierarchy of Poisson tensor fields
$\cal  P_k:=\cal  N_G^k\, \cal E\  , k\in {\Bbb Z}$.
However these tensor fields are not invariant along the flow of the Garnier
vector
field $\cal  Y_G $ \eqref{2gGY1}, so they do not generate  a bi--Hamiltonian
hierarchy starting  from $\cal  Y_G$.

\subsection{A new integrability criterion}
\label{SSect. TeoC}
 In the previous subsection  we have put into
evidence  some  problems arising  in the geometrical reduction of a
bi--Hamiltonian structure from an extended phase space onto the original one.
  As an alternative construction, here we  introduce a new
integrability scheme, weaker than the   bi--Hamiltonian one, but living in the
standard phase
space. We shall define this new structure for a generic Hamiltonian system with
$n$ degrees of freedom;  for $n=2$ it coincides with the one introduced in
\cite{francesi} for the
H\'enon--Heiles system with the Hamiltonian \eqref{2gfHham} and $a_4=0$. As
new examples of this integrability structure,  the case of the Garnier system
with two
degrees of freedom will be discussed here whereas multidimensional extensions
of the
H\'enon--Heiles system will be presented elsewhere.

\begin{prop} \label{TeoC}
Let  $M$ be a  $2n$--dimensional Poisson manifold equipped with  a Poisson
tensor
$Q_0$, and  $Z_0$  a Hamiltonian vector field
 with Hamiltonian $h_0$: $Z_0=Q_0 \, dh_0$.
Let there exist
a tensor  ${\cal  N}:TM\rightarrow TM$ and a skew--symmetric tensor
$Q_1:T^*M\rightarrow TM$   such that
\begin{equation} \label{ipQ1=NQ0}
Q_1={\cal  N} Q_0 \ .
\end{equation}
 Denote by $Z_i$ and $\alpha_{i}$ the vector fields   and the $1$--forms
obtained,
respectively, by the
iterated action of the tensor  ${\cal  N}$ on  $Z_0$ and its adjoint
 ${\cal  N}^*: T^*M\rightarrow T^*M$   on $\alpha_0:=dh_0$

\begin{equation}
Z_i:={\cal  N}^i Z_0 \ , \qquad \alpha_{i}:={\cal  N}^{*^i}\alpha_0 \qquad
(i=1,\ldots,n-1) \ .
\end{equation}
\par
Let  there exist $n-1$ independent functions $h_i\  (i=1,\ldots ,n-1)$ and
$(n^2+n-2)/2$ functions $\mu_{ij} \  (i=1,\ldots,n-1; 0\leq j \leq i)$ with
$\mu_{00}=1$,
$\mu_{ii}\neq 0 \  (i=1,\ldots,n-1)$, such that the 1-forms
$\alpha_{i}$  can be written as

\begin{equation} \label{ipalpha=mudh}
\alpha_i =\sum_{j=0}^i \mu_{ij}\, dh_j\qquad (i=1,\ldots,n-1)\  .
\end{equation}
Under the previous assumptions   the following results hold:
\par
\par\noindent
{\bf i)}
the vector fields $Z_i$ satisfy the recursion relations

\begin{equation}  \label{ZbiQ}
Z_{i+1}=Q_0\alpha _{i+1}=Q_1\alpha _{i} \qquad \qquad (i=0,\ldots, n-2) .
\end{equation}
\par
\par\noindent
{\bf ii)}
the functions $h_i$ are in involution with respect to the Poisson bracket
defined by $Q_0$
 and they are constants of motion for the fields $Z_k$

\begin{equation} \label{invh}
\{h_i, h_j\}_{Q_0}=0 \ ,  \qquad\qquad \cal L_{Z_k}(h_i)=0  \ ,
\end{equation}
where $\cal L_{Z_k}$ denotes the Lie derivative with respect to the vector
field $Z_k$.
\par
\par\noindent
{\bf iii)}
the Hamiltonian system corresponding to the vector field $Z_0$  is
Liouville--integrable.
In addition if $Q_1$  is a Poisson tensor field, then also $Z_1$  is an
integrable
Hamiltonian vector field and the functions $h_i$ are in involution also with
respect to the
Poisson bracket defined   by $Q_1$.
\end{prop}
\begin{pf} \hfill
\par\noindent
{\bf i)}
{}From Eq. \eqref{ipQ1=NQ0} and the skew--symmetry of $Q_0$ and $Q_1$ it
follows that
$Q_0{\cal  N}^*={\cal  N}Q_0$ and $ Q_1{\cal  N}^*={\cal  N}Q_1$. Then

\begin{equation}
Z_1-Q_0\alpha_1=Z_1-Q_0 {\cal  N}^*\alpha_0=Z_1-{\cal  N}Q_0\alpha_0=0
\end{equation}
and the first relation \eqref{ZbiQ} is proved by induction since it is

\begin{equation}
Z_{i+1}-Q_0\alpha_{i+1}={\cal  N} Z_i-Q_0 {\cal  N}^*\alpha_i={\cal  N}
(Z_i-Q_0\alpha_i) \ .
\end{equation}

The second relation \eqref{ZbiQ} follows from

\begin{equation}
Z_{i+1}-Q_1 \alpha_i={\cal  N}Z_i-Q_1\alpha_i={\cal  N}(Z_i-Q_0\alpha_i)\  .
\end{equation}
\par
\par\noindent
{\bf ii)}
By \eqref{ipalpha=mudh},  the gradients $dh_k$ can be expressed for any $k$ in
terms of
$dh_0$

\begin{equation}
dh_k= (\sum_{i=0}^k \nu_{ki} {\cal  N}^{*^i} )dh_0 \ ,
\end{equation}
where  $\nu_{ki}$ are the elements of the  matrix $a^{-1}$, $a$ being
 the lower triangular matrix defined by $a_{ij}=\mu_{ij}\, (i\geq j),\; a_{ij}
=0\,  (i < j), \;
(i,j=0,\ldots,n-1)$. Thus

\begin{equation}
\begin{split}
\{h_i, h_j\}_{\scriptscriptstyle{Q_0}}:&= \langle dh_i, Q_0 dh_j\rangle  \\
                            &=\sum_{a=0}^i\sum_{b=0}^j \nu_{ia}\nu_{jb}
                                                     \langle {\cal
N}^{*^a}dh_0,Q_0{\cal N}^{*^b}dh_0\rangle \\
                            &= \sum_{a=0}^i\sum_{b=0}^j \nu_{ia}\nu_{jb}
                                                               \langle dh_0,
{\cal  N}^{a+b}Q_0 dh_0\rangle
\end{split}
\end{equation}
and the first relation \eqref{invh} follows from  the skew--symmetry of the
tensor ${\cal  N}^m
Q_0$ for any $m$.  Furthermore

\begin{equation}
\begin{split}
\cal L_{Z_k}(h_i) &=\langle dh_i, Q_0 \alpha_{k-1}\rangle \\
                    &=\langle dh_i, Q_0 \sum_{j=0}^k \mu_{kj} dh_j \rangle \\
                    &= \sum_{j=0}^k
\mu_{kj}\{h_i,h_j\}_{\scriptscriptstyle{Q_0}} \\
                    &=0
\end{split}
\end{equation}
\par
\par\noindent
{\bf iii)}
Since $Z_0$ is a Hamiltonian vector field, it is Liouville--integrable on
account of the previous
result. Moreover, since it is

 \begin{equation}
\begin{split}
\{h_i, h_j\}_{\scriptscriptstyle{Q_1}}:&= \langle dh_i, Q_1 dh_j\rangle  \\
                            &=\sum_{a=0}^i\sum_{b=0}^j \nu_{ia}\nu_{jb}
                                                               \langle {\cal
N}^{*^a}dh_0,Q_1{\cal  N}^{*^b}dh_0\rangle
\\
                            &= \sum_{a=0}^i\sum_{b=0}^j \nu_{ia}\nu_{jb}
                                                               \langle
dh_0,{\cal  N}^{a+b}Q_1 dh_0\rangle \\
                             &=0 .
\end{split}
\end{equation}
it follows that if  $Q_1$ is also a Poisson tensor,  $\{ , \}_{Q_1}$ is a
Poisson bracket, $Z_1$ is a
Hamiltonian  vector field and then it  is Liouville--integrable.
\end{pf}

\begin{rem}
  The recursion scheme and the integrability of the vector field $Z_0$ do not
require
that the skew--symmetric tensor $Q_1$ be a Poisson tensor; so $M$ is a Poisson
manifold,
not a bi-Hamiltonian one.
$\square$
\end{rem}
In view of the  applications of the next subsection, it may be worthwhile to
remark that the
results of  Prop. \ref{TeoC} hold true if the role of $Q_0$ and $Q_1$ are
interchanged; to be more
precise, one can prove (just as for Prop. \ref{TeoC} )

\begin{prop}
The integrability scheme of Prop. \ref{TeoC}  is still valid if $Q_0$ is
skew--symmetric, $Q_1$ is
a Poisson tensor and the role of $Z_0$ is now played by $Z_1=Q_1dh_0$. The
involution relations
\eqref{invh} become $\{h_i, h_j\}_{Q_1}=0$.
\end{prop}

\subsection{The integrability structure of  the H\'enon--Heiles and  the
Garnier systems}
\label{SSect2gHGintstr}

In Subsect. \ref{SSect2gHredstr} we have recovered by projection onto the
quotient
manifold   $S_2$  the Nijenhuis tensor  \eqref{2gHN}
and a hierarchy of compatible Poisson tensors; however, it is not possible to
associate to these tensors and to the H\'enon--Heiles vector field $X_1$
\eqref{2gHX0}
a bi--Hamiltonian hierarchy of vector fields. Nevertheless it is possible to
use these
elements to construct  an example of the integrability structure
introduced in Prop. 5.2. To this purpose, let us make the following choices:

\begin{itemize}
\item[i)]
 $Q_1=E$, the vector field $Z_1:=X_1$ \eqref{2gHX0} with Hamiltonian $h_0:=H_0$
\eqref{2gHham};
\item[ii)]
the tensor field ${\cal  N}:=N_H$  \eqref{2gHN} and $Q_0:=P_{-2}=N_H^{-2}P_H$,
with
$P_H$ as in \eqref{2gHP0};
\item[iii)]
the function $h_1:=H_1$ \eqref{H2H} and the functions $\mu_{ij}$ as
$\mu_{10}=0,
\mu_{11}=1/q_2^2$;
\end{itemize}
then it is immediate to check that the conditions of   Prop. 5.2 are satisfied.
Moreover the vector
field
$Z_0:=Q_0\, dh_0=P_{-2}\, dH_0$ is a new integrable vector field:

\begin{equation}
Z_0=
\begin{bmatrix}
-2p_1q_1-p_2q_2 \\\
-p_1q_2 \\\
-p_2^2+6q_1^3+2q_1q_2-\frac{a_4}{4q_2^2}-\frac{a_1}{2}q_1-2a_0q_1^2+\frac{a_0}{4}q_2^2
\\ \
p_1p_2+\frac{q_2^3}{2}+3q_1^2q_2-\frac{a_1}{4}q_2+a_0q_1q_2
\end{bmatrix}  \ .
\end{equation}
This integrability structure is related,  through the map
\eqref{mapHHf}, to  the one introduced
in \cite{francesi} for the Hamiltonian \eqref{2gfHham} with $a_4=0$.
\par
For the Garnier system with two degrees of freedom
    one can construct an example of the
integrability structures of Prop. \ref{TeoC}. Indeed   if one  uses the
elements of
Subsect. 5.1 and  makes  the following choices:

\begin{itemize}

\item[i)]
$ Q_0:=\cal  E$,  $ h_0:=\Tilde{\cal  G}_1$ \eqref{2gGG} ,
$ Z_0:=\cal  Y_G$  \eqref{2gGY1} ;

\item[ii)]
 ${\cal  N}:={\cal  N}_G^{-1}=\cal E\, \cal P_G^{-1}$,  with $\cal P_G$ as in
(5.3) ,
$\cal  Q_1:=\cal  P_{-1}={\cal  N}_G^{-1} \, \cal E$ ;

\item[iii)]
the functions $ h_1:=\Tilde{\cal  G}_2$ \eqref{2gGG} ,   $\mu_{10}=0$,
$\mu_{11}=-\frac{\lambda_1^2\lambda_2^2}
{\lambda_2\psi_1^2+\lambda_1\psi_2^2-\lambda_1\lambda_2}$;
\end{itemize}
then the conditions of   Prop. \ref{TeoC} are satisfied. Moreover the vector
field
$Z_1:=\mu_{11}\cal  Y_2$  is a new integrable vector field ( $\cal  Y_2$ is the
 restriction to the
submanifold of $\cal M_2$,  $d=cost$,   of the vector field
$\Tilde{\cal Y}_2 $ \eqref{2gGY}).

At last,  we compute how    the map between the standard phase spaces of the
H\'enon--Heiles
 and  of the Garnier systems, induced by the map  \eqref{map2HG}, acts on the
recursion
operators of the previous integrability structures.

\begin{prop}
Let us consider the map   $\Psi: (q_1,q_2,p_1,p_2)\mapsto
(\psi_1,\psi_2,\chi_1,\chi_2)$

\begin{equation}
\begin{split}   \label{rmap2HG}
\psi_{1} &= \lambda_{12}^{-1/2}(\lambda_{1}^2 +2\lambda_{1}q_{1}
-q_{2}^2)^{1/2}
 \ ,\qquad\qquad
\psi_{2} =( \lambda_{12})^{-1/2} (-\lambda_{2}^2-2\lambda_{2}q_{1}
+q_{2}^2)^{1/2} \ , \\
\chi_{1}& = \frac {\left( \lambda_{1}p_{1} -q_{2}p_{2}\right)}
{\big(\lambda_{12}\ (\lambda_{1}^2
                 +2\lambda_{1}q_{1} -q_{2}^2)\big)^{1/2}} \ , \qquad
\chi_{2} = \frac {\left( \lambda_{2}p_{1} -q_{2}p_{2}\right)}
{\big(\lambda_{12}\ (-\lambda_{2}^2
                 -2\lambda_{2}q_{1} +q_{2}^2)\big)^{1/2}}  \ .
\end{split}
\end{equation}
The map $\Psi$ relates the recursion operators of the H\'enon--Heiles and of
the Garnier systems:
$\Psi_*  N_H  =\cal N_G^{-1}\Psi_* $.

\end{prop}

\section{Concluding remarks}
In this paper we have derived a bi--Hamiltonian formulation for stationary
flows,
and for the first restricted flows of the KdV hierarchy.  Our approach amounts
to searching
the kernel of, respectively, the Poisson pencil and
$n$--Poisson structures extracted from the Poisson pencil of the KdV hierarchy.
In this
approach the  generating function of the GD polynomials plays a relevant role.
Moreover it
allows us to construct a map between stationary flows and restricted flows;  in
the case
of the fifth-order stationary KdV equation, this map relates  solutions of the
H\'enon--Heiles
system with solutions of the Garnier system.
 However,  to obtain these results one must extend
 the phase space of the reduced flows by means of some free
parameters naturally contained in the corresponding Hamiltonian functions. This
difficulty can
be overcome, at least if one analyzes complete integrability of a Hamiltonian
system without
requiring an explicit knowledge of a bi--Hamiltonian structure. To this
purpose, we have
introduced a new integrability scheme in the standard phase space, which
implies  Liouville
integrability of the reduced Hamiltonian systems.
For brevity we have applied this scheme only to the H\'enon--Heiles and the
Garnier systems
with two degrees of freedom. Other examples such as H\'enon--Heiles type
systems with
three and four degrees of freedom, constructed by means of the  reduction
method of
Sect. 2, will be discussed elsewhere.
\par

\vspace{1truecm}
{\bf Acknowledgments.}
I  wish to thank F. Magri, who pointed out the role of the GD polynomials in
the
bi--Hamiltonian formulation of the KdV hierarchy and C. Morosi for many
valuable
discussions and suggestions.
\vfill
\pagebreak

\end{document}